\documentclass[aps,prl,preprint,superscriptaddress]{revtex4}
\usepackage{graphicx}
\usepackage{morefloats}
\usepackage{color}
\usepackage{amssymb}
\usepackage{adjustbox}

\usepackage[english]{babel}
\usepackage{amsmath}

\begin{document}


\title{Observation of magnetic skyrmions in unpatterned symmetric multilayers at room temperature and zero magnetic field}

\author{J. Brand\~{a}o}
\affiliation{%
Laborat\'{o}rio Nacional de Luz S\'{i}ncrotron, Centro Nacional de Pesquisa em Energia e Materiais, 13083-970 Campinas SP, Brazil}%

\author{D. A. Dugato}
\affiliation{%
Laborat\'{o}rio Nacional de Luz S\'{i}ncrotron, Centro Nacional de Pesquisa em Energia e Materiais, 13083-970 Campinas SP, Brazil}%
\affiliation{%
Departamento de F\'{i}sica, Universidade Federal de Santa Maria, 97105-900 Santa Maria RS, Brazil}%

\author{R. L. Seeger}
\affiliation{%
Departamento de F\'{i}sica, Universidade Federal de Santa Maria, 97105-900 Santa Maria RS, Brazil}%

\author{J. C. Denardin}
\affiliation{%
Departamento de F\'{i}sica, Universidade Federal de Santa Maria, 97105-900 Santa Maria RS, Brazil}%
\affiliation{%
Departamento de F\'{i}sica and CEDENNA, Universidad de Santiago de Chile, 9170124 Santiago, Chile}%

\author{T. J. A. Mori}
\affiliation{%
Laborat\'{o}rio Nacional de Luz S\'{i}ncrotron, Centro Nacional de Pesquisa em Energia e Materiais, 13083-970 Campinas SP, Brazil}%

\author{J. C. Cezar*}
\affiliation{%
Laborat\'{o}rio Nacional de Luz S\'{i}ncrotron, Centro Nacional de Pesquisa em Energia e Materiais, 13083-970 Campinas SP, Brazil}%

\date{\today}

\begin{abstract}

Magnetic skyrmions are promising candidates  for the next generation of spintronic devices  due to  their small size and  topologically protected structure. One challenge for using these magnetic states in   applications  lies on controlling the  nucleation process and  stabilization that usually requires an external force. Here, we report on the evidence of skyrmions in unpatterned symmetric Pd/Co/Pd multilayers at room temperature without prior application of neither electric current nor magnetic field. By decreasing the ferromagnetic interlayer thickness, the tuning of the physical properties across the ferromagnetic/non-magnetic interface gives rise to a transition from worm like domains patterns to isolated skyrmions as demonstrated by magnetic force microscopy.  On the direct comparison of the measured and simulated skyrmions size, the quantitative interfacial Dzyaloshinskii-Moriya interaction (iDMI) was estimated, reveling that isolated skyrmions are just  stabilized at zero magnetic field for low values of iDMI.  Our findings provide new insights towards the use of stabilized skyrmions for room temperature devices in nominally symmetric multilayers.


\end{abstract}

\maketitle

\begin{figure*}
\centering
\includegraphics[width=\linewidth]{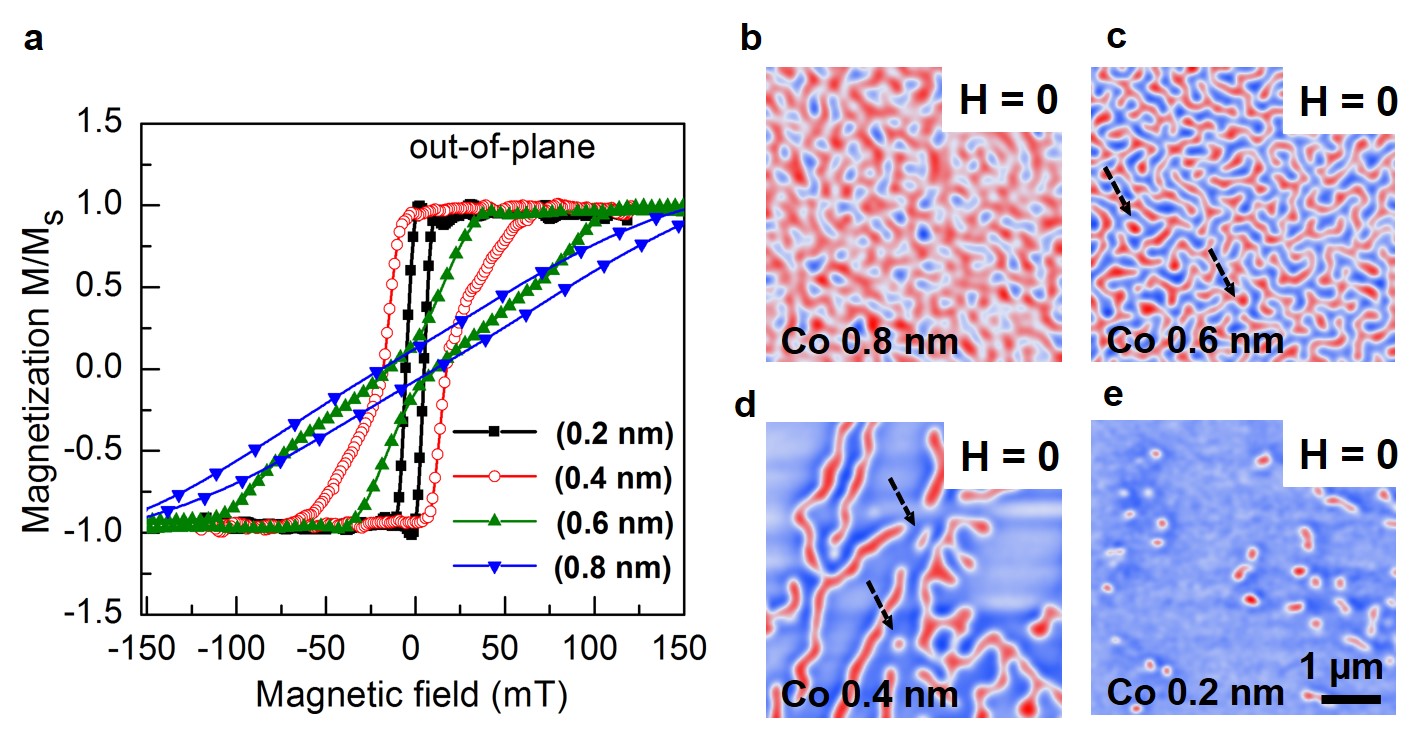}
\caption{{ Magnetization curves and MFM images. a, hysteresis curves evolve from tail-like to square loop for decreasing Co thicknesses. {\bf b-e,} The MFM images acquired for different Co thicknesses in the as-gown state revealing the transition from small up and down domains  b-c, to isolated skyrmions  e. Red and blue contrasts represent the magnetization out-of-plane of the film.}
\label{brandao1a}}
\end{figure*}

The full control of the magnetization processes in small-sized magnetic states is essential  to increase the density of information in magnetic memories \cite{jeovani1,jeovani2,jeovani6}. Magnetic skyrmions, i.e. nanometer-sized topological defects presenting swirling spin textures, are promising candidates for achieving efficiency and functionality towards room temperature devices \cite{review,jeovani8,jeovani9,jeovani9a,jeovani9c}. Currently, the standard heterostructures for studying skyrmions are heavy metal/ferromagnetic (HM/FM) interfacial systems, with structural inversion asymmetry in ultrathin films fabricated  by magnetron sputtering. The strong spin orbit coupling (SOC) of the HM layer can lead to an antisymmetric exchange known as interfacial Dzyaloshinskii-Moriya interaction (iDMI) \cite{jeovani19, jeovani20}, which plays a key role in the stabilization of chiral spins textures such as skyrmions \cite{jeovani21,jeovani22,jeovani7}. The iDMI emerges in HM/FM interfaces owing broken spatial inversion symmetry which determines its sign and direction, whilst the iDMI magnitude depends on the SOC \cite{jeovani23, jeovani24,jeovani25}. In this sense, different combinations of HM/FM interfaces have been investigated to obtain strong iDMI amplitudes  and distinct signs to stabilize skyrmions and define their chirality \cite{jeovani26,jeovani27,chiral}. Although isolated skyrmions have been observed recently at room temperature  in  this kind of systems, their nucleation and stabilization in most cases require external magnetic field and/or electrical current \cite{jeovani28,jeovani29,jeovani30}. It would be interesting also to nucleate skyrmions without the need of any external force (magnetic field or current). This was obtained using lithographically made structures to confine single or multiple skyrmions depending on the geometric parameters of the patterned samples \cite{jeovani31,jeovani32}. It remains to be demonstrated that one can observe the spontaneous formation of skyrmions, even without nanostructured confinement.

In this letter, we demonstrate that magnetic skyrmions can be stabilized at room temperature in unpatterned samples without the need for any preceding external excitation nor geometric confinement. We observe the formation of  skyrmions  in both as-grown (not exposed to the magnetic field) and remnant states of nominally symmetric Pd/Co/Pd multilayers when the Co thickness is as thin as  the percolation threshold  of a continuous layer. The tuning of the magnetic properties at the HM/FM interfaces by simply thinning the FM layer leads to a transition of the magnetic domains pattern from a labyrinthine state, passing through a state with long and separated stripes, then reaching isolated skyrmions  for the lowest Co thickness. The  observation of skyrmions  is investigated mainly by magnetic force microscopy (MFM). We verify their reproducibility and density along the unpatterned thin film by imaging different regions of the sample to obtain the skyrmions average size. Besides,  magnetization curves show the dependence of the perpendicular magnetic anisotropy (PMA) and remnant magnetization with the Co thickness, providing a direct link to understand  the magnetic textures observed in the  MFM images. Micromagnetic simulations are also carried out to elucidate the role of each magnetic parameter  upon the  skyrmions stability at room temperature and zero external magnetic field.


We  prepared twin samples in the same sputtering runs (see methods), in order to take MFM images at zero magnetic field and  to measure hysteresis loops. The out-of-plane magnetization curves (normalized to the saturation magnetization M$_s$) are summarized in  Fig. \ref{brandao1a}a.  For  thicker Co samples (0.8 and 0.6 nm) the hysteresis show a tail feature, whilst thinner  (0.4 and 0.2 nm) present a more square-shaped format. In-plane hysteresis loops were also performed (see supplementary information S1). 

To image the magnetic domains patterns without any prior applied magnetic field, MFM images were firstly acquired in samples in the as-grown state. For  Co (0.8 and 0.6 nm), the MFM images show that the magnetization is broken in small domains (Fig. \ref{brandao1a}b-c). More specifically, for Co (0.6 nm) the magnetic domains exhibit a clear worm-like configuration in the so-called labyrinthine state  (Fig. \ref{brandao1a}c).  Reducing the Co thickness to 0.4 nm, the magnetic domain pattern shows long  and separated stripes (Fig. \ref{brandao1a}d). It is very interesting that some  skyrmions are clearly observed among both the worm-like and the long stripes patterns discussed so far, see the indications by dashed black arrows in  Fig. \ref{brandao1a}c-d. In a previous work, similar skyrmions among the worm-like domain patterns were  observed by MFM images \cite{jeovani35}.

\begin{figure*}
\centering

\includegraphics[width=1.1\linewidth]{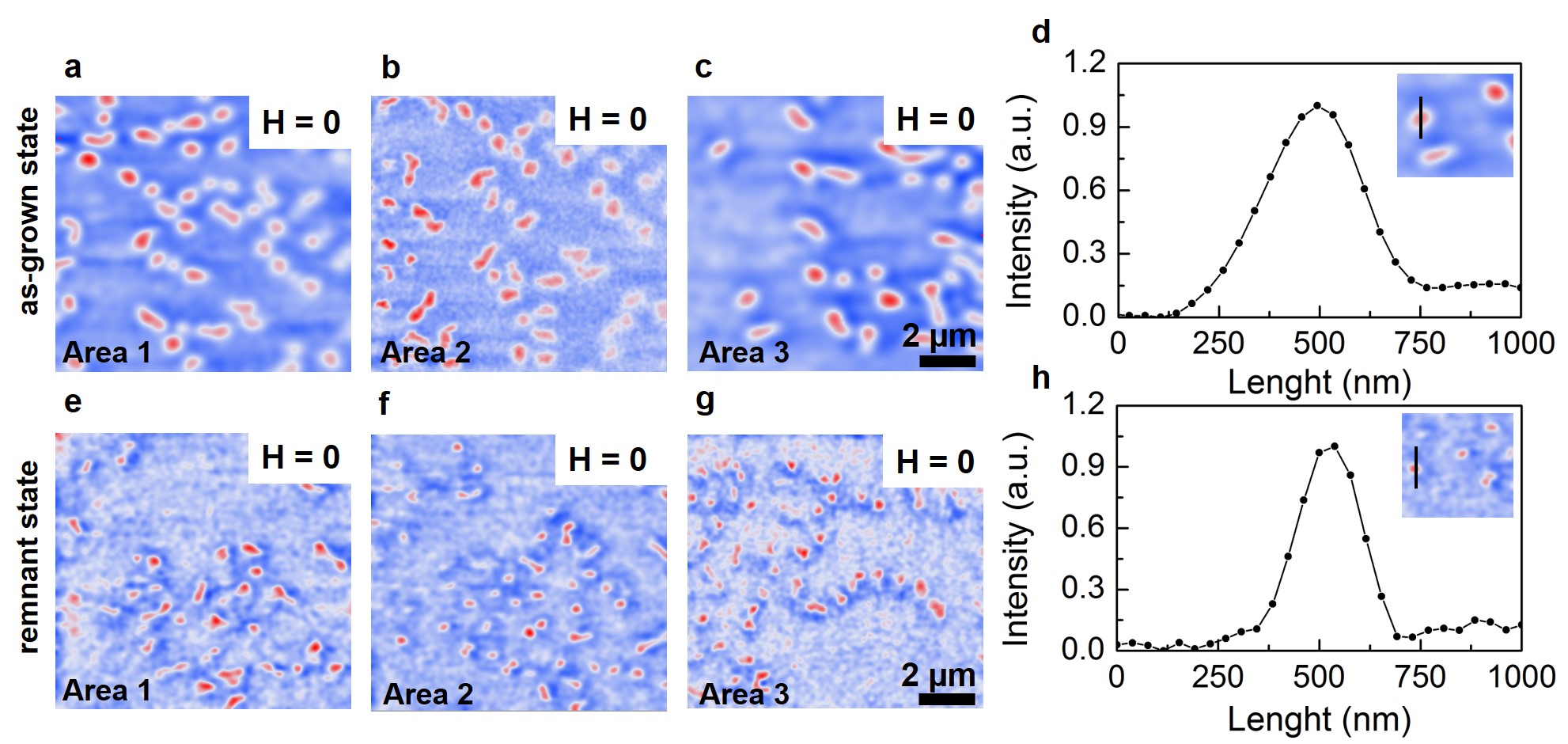}
\caption{{ Representative MFM images of isolated skyrmions. a-c, in the as-grown state.  {\bf e-g,} in the remnant state.  Isolated skyrmions are observed at zero magnetic field over different places on the sample surface.  d-h, the profile shows the shape of the isolated skyrmions. Most notable, the skyrmions size is larger in the as-grown than the remnant state. The MFM images obtained in different locations highlight the homogeneity on the skyrmions formation.}
\label{Brandao2a1}}
\end{figure*}

By further thinning  the Co layer, the transition from worm-like to long stripes  evolves to a new magnetic domain pattern for the thinnest Co (0.2 nm). As it is shown in  Fig. \ref{brandao1a}e, many isolated skyrmions are observed. These images reveal that no prior stabilizing field or injection current is required to generate isolated skyrmions. 

The MFM image shown in Fig. \ref{brandao1a}e was acquired for Co (0.2 nm) sample which has never submitted to external magnetic field. However, the hysteresis loops measured for one identical sample, can be used to explain this unexpected  behavior.  The Co (0.2 nm) is the sample whose out-of-plane hysteresis loop is the most square-shaped and which presents the highest remnant state and PMA (Fig. \ref{brandao1a}a). Thus, at zero magnetic field, owing high remanence and PMA, the magnetic pattern would be more favorable  oriented out-of-plane in a quasi-uniform magnetization state. Instead, in our MFM images, skyrmions appear before reaching this ground state, suggesting that carefully controlling  the transition from worm-like to single domain varying the Co interlayer thickness, it is feasible to create isolated stabilized skyrmions. Therefore, this combination of PMA and remnant magnetization emerges as one of the main sources for the generation of skyrmions at room temperature and zero magnetic field in Pd/Co/Pd symmetric multilayers. 

On the other hand, it is well known that  skyrmions are stabilized by the presence of interfacial Dzyaloshinskii-Moriya interaction (iDMI) \cite{jeovani26}, which occurs in systems with structural inversion asymmetry \cite{jeovani39,jeovani40}. Here, ideal symmetric Pd/Co/Pd multilayers should not have structural inversion asymmetry, so  both Co/Pd and Pd/Co interfaces should contribute with iDMIs of same amplitude but opposite signs, hence leading to a null liquid iDMI strength. However, it was observed that crystallographic asymmetry between Pt/Co and Co/Pt interfaces in Pt/Co/Pt systems can give rise to liquid iDMI \cite{jeovani41}. Also, the total magnetic moment induced in the Pd is larger at the top Co/Pd interface than at the bottom Pd/Co, leading to an asymmetric magnetic proximity effect in Pd/Co/Pd trilayers \cite{jeovani42}. This behavior may also contribute to iDMI in symmetric HM/FM/HM systems as the proximity effect might have a correlation with iDMI as  suggested in the ref. \cite{jeovani43}. Furthermore, iDMI has been confirmed as the origin of asymmetric domain wall creep measured in symmetric Pd/Co/Pd multilayers by means of polar Kerr images \cite{jeovani34}. Supported by these considerations and  references, we consider that our symmetric Pd/Co/Pd multilayers may also present weak iDMI. Indeed, we will show through the correlation between micromagnetic simulations and MFM images that isolated skyrmions at zero magnetic field are just stabilized in symmetric multilayers  considering a weak but no-null iDMI.


A question that emerges here is whether the transition from worm-like  to isolated skyrmions can be observed over the entire unpatterned sample. To verify this point, several MFM images were acquired on different regions of each sample (see methods). To illustrate, 3 different images are shown in  Fig. \ref{Brandao2a1}a-c, all of them exhibit many skyrmions, strengthening that their spontaneous formation is not an isolated case but rather is reproducible over different areas of the sample surface. 

So far we demonstrated the observation of magnetic skyrmions at room temperature without any prior applied magnetic field. In order to be useful for applications, the skyrmions should survive also after being subjected to an external magnetic field. The samples were then imaged by MFM in the remnant state after the measurements of the hysteresis loops shown in Fig. \ref{brandao1a}a. The images of the samples Co (0.8 nm,0.6 and 0.4 nm), see (supplementary information S2), present worm-like and long stripe patterns with a few isolated skyrmions, being  very similar to the ones  observed for the as-grown samples. The same tendency of similar magnetic features was observed for the thinnest Co (0.2 nm), whose isolated skyrmions  can be visualized in Fig. \ref{Brandao2a1}e-g. Similarly to the as-grown state images, these 3 MFM measurements are representative of the magnetic domains patterns over the entire unpatterned sample and proves that skyrmions can be generated in symmetric Pd/Co/Pd multilayers equally in its remnant state after magnetization cycles. Notwithstanding, it is worth noting that the isolated skyrmions presented in the Fig. \ref{Brandao2a1}  are randomly distributed and their size and shape are not identical. We suggest that this distribution arises from minor structural and morphological inhomogeneities along the different regions of the film as the Co layers are ultrathin. Besides, the variation of magnetic parameters such as $M_{s}$, PMA, and iDMI also contributes to the different size and shape of stabilized skyrmions \cite{skyrmions200nm}.

\begin{figure*}
\centering
\includegraphics[width=\linewidth]{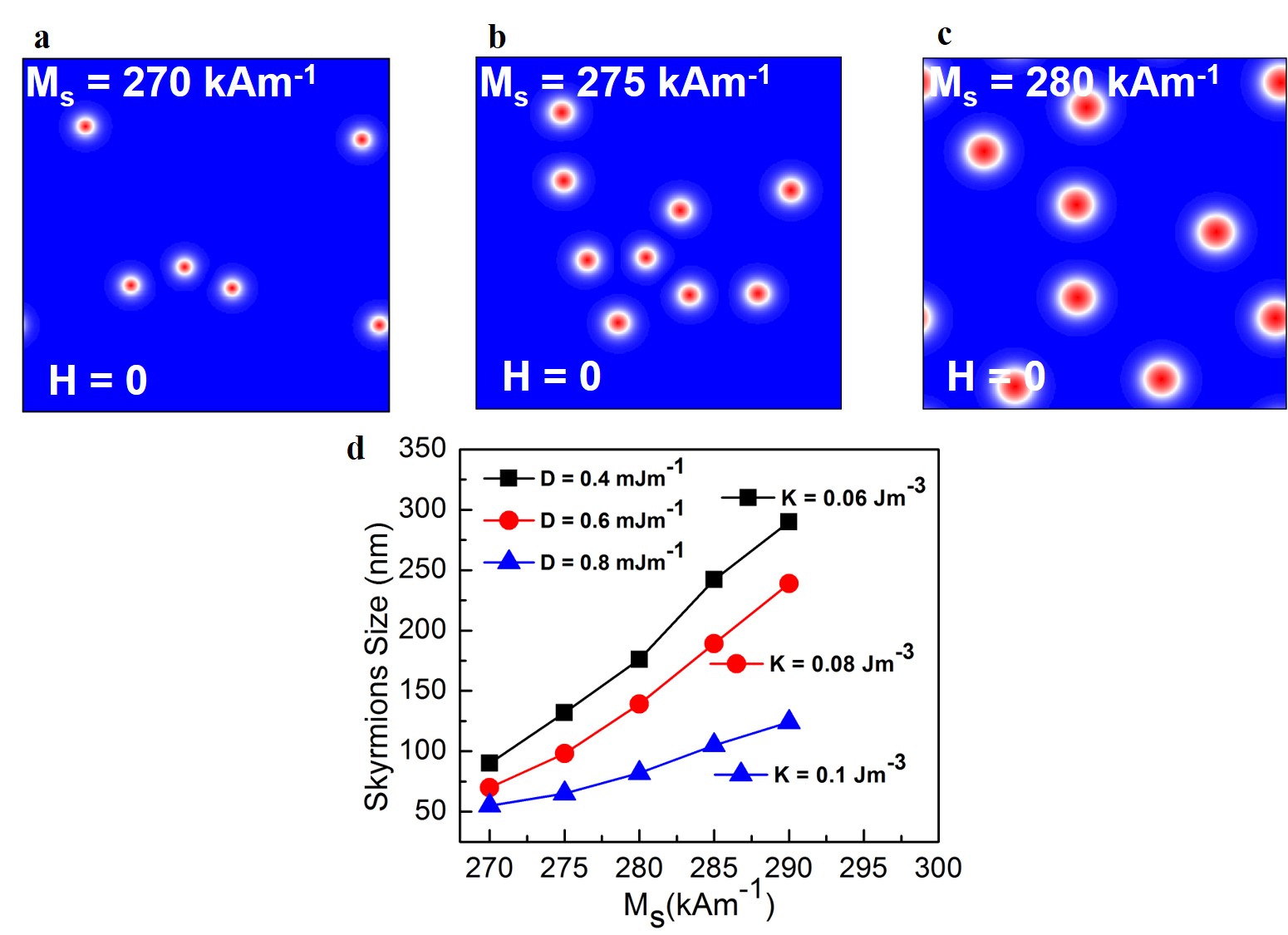}
\caption{{ Simulated stabilized skyrmions at zero magnetic field for $D$ = 0.4 mJm$^{-1}$. a-c, the skyrmions are isolated and randomly distributed as it was observed experimentally. The saturation magnetization parameter was varied showing that the  skyrmions becomes larger as $M_s$ increases. In d, the skyrmions size extracted from the simulations to different $D$ and $k$. The skyrmions is larger when $k$ and $D$ are reduced for a ranging of $M_s$  values.} 
\label{Brandao4}}
\end{figure*}

Exploring  the  formation of skyrmions for  the Co (0.2 nm) sample, we obtained their average size and density distribution for both as-grown and remnant states analyzing  10$\times$10 $\mu$m$^2$ MFM images from 10 different regions on the film. We used the full width at half maximum (FWHM) \cite{size}, extracted from the line scan of 5 isolated skyrmion at each image to determine its size, as it is illustrated in the Fig. \ref{Brandao2a1}d-h for an isolated skyrmion in the as-grown and other in the remnant state, respectively. The as-grown skyrmions present an average size of 314 nm with a standard deviation of  41 nm, and are distributed with an average density  of $\sim$ 0.3 skyrmions  per $\mu$m$^2$. The skyrmions detected in the remnant state, in  turn, exhibit a reduced average size  of $\sim$  165 nm with a standard deviation  of 32 nm and a density of 0.25 skyrmions per $\mu$m$^2$. 

The difference  in the skyrmions size and density after submitting  the sample to magnetic field represents a reduction of $\sim$ 55\% and $\sim$ 18\%, respectively. This comes from the fact that the remanence is larger than the magnetization in as-grown state, what leads to smaller size and density of the skyrmions.  Nonetheless, this observation demonstrates that skyrmions in these samples are homogeneous and robust, persisting even after the application of an external magnetic field. These skyrmions observed in larger areas  in combination with patterned nanostructures can be used as a source to feed constrictions in technological applications \cite{large}. The advantage would be the fact that they do not need any stabilizing field to transform magnetic domains in a worm like configuration into skyrmions. Therefore, they can be manipulated with ultra-low density current which is beneficial to perform low-energy spintronics devices.



In order to understand the formation of stabilized skyrmions at zero magnetic field, micromagnetic simulations  were carried out  using Mumax$^3$ code \cite{Mumax} (see methods). The physical parameter extracted from the hysteresis loops was the  magnetization saturation $M_s$. More specifically, for Co 0.2 nm, the measured $M_s$ was $\sim$ 280 kA/m. The value is consistent with previous reported on Pd/Co multilayers \cite{jeovani45,jeovani46,jeovani47}. We varied the magnetization saturation $M_s$ and magnetic anisotropy $k$ to understand the impact of these properties on the  skyrmions stability considering a weak Dzyaloshinskii-Moriya below $D$ = 0.8 mJm$^{-2}$. 

We start by showing examples of micromagnetic simulations on the stabilized skyrmions at zero magnetic field considering a constant $D$ = 0.4 mJm$^{-2}$. Fig. \ref{Brandao4} a-c  shows  isolated skyrmions when $k$ is 0.06 MJm$^{-3}$ and $M_{s}$ ranging  from 270 kAm$^{-1}$ to 280 kAm$^{-1}$. Most notable, the higher the $M_{s}$ the  larger  the skyrmions size, suggesting that  local variations in the magnetic properties ($k$, $M_{s}$ and $D$) beyond the non-uniformity interfaces play also a  role in modifying the skyrmions size. Indeed, this behavior is summarized in Fig. \ref{Brandao4} d. Varying $M_{s}$, we obtained simulated skyrmions size ranging from $\sim$ 50 nm to  $\sim$ 290 nm considering different values of $D$ and $k$. It indicates that larger skyrmions are stabilized when both $D$ and $k$ are reduced while $M_{s}$ is increased. Comparing these results with the skyrmions size measured in the remnant and as grown state, it represents a good agreement. This direct comparison between simulation  and experimental on the skyrmions size,  allowed us quantitatively estimate the iDMI in our sample as  $D$ = 0.4 - 0.6 mJm$^{-2}$.  Similar results were obtained by performing simulations using negative $D$ values, see (supplementary information S3). 




\begin{figure*}
\centering
\includegraphics[width=1.05\linewidth]{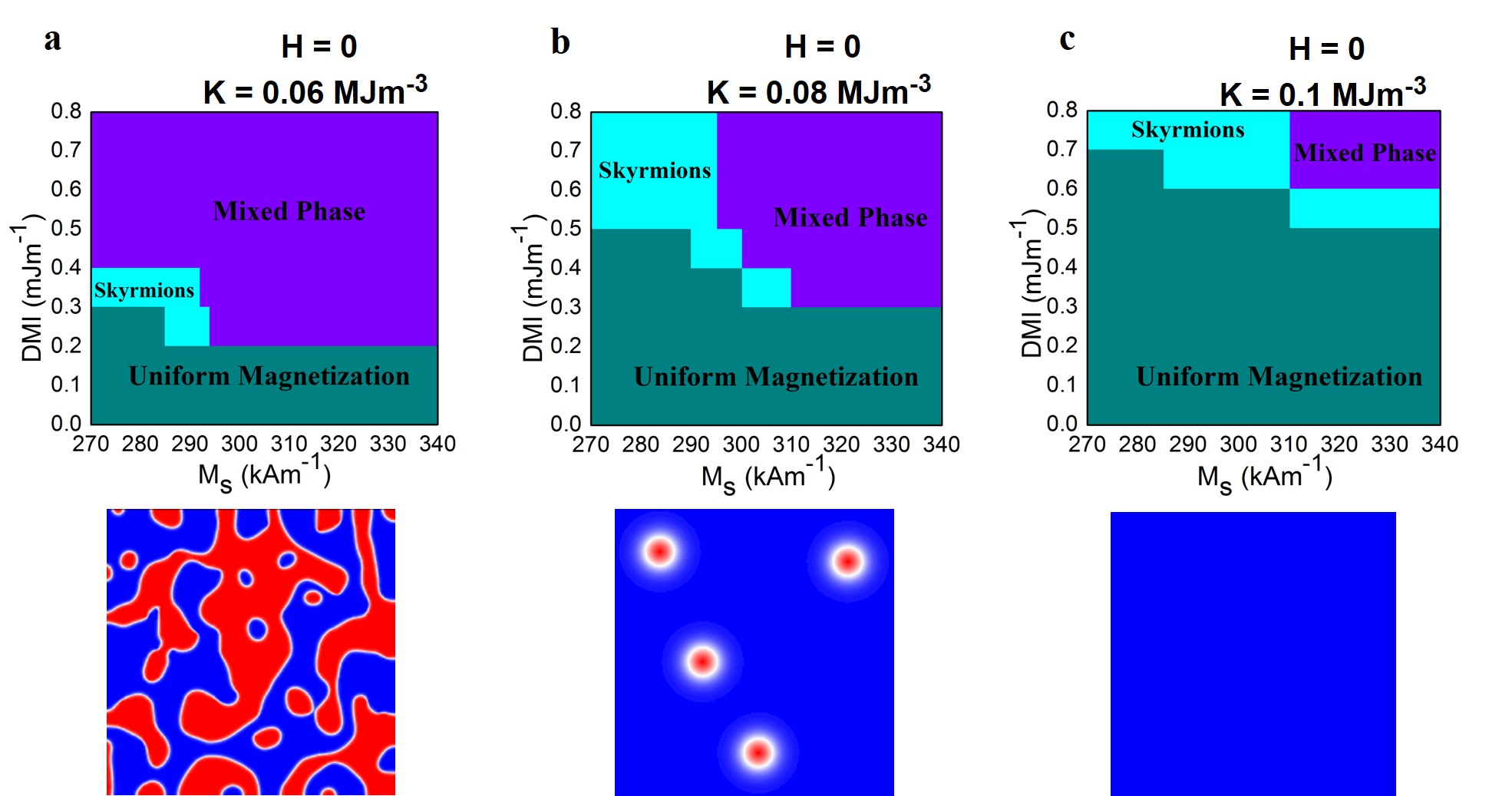}
\caption{{Simulated phase diagram. The ground state was determined by modifying $D$ and $M_s$ for different $k$. a, below $D$ = 0.2 mJm$^{-1}$ the uniform magnetization pattern is the  ground state. Above $D$ = 0.2 mJm$^{-1}$ isolated skyrmions are stabilized for $M_s$ ranging from 270 - 295 Am$^{-1}$ and  a mixed phase (skyrmions and stripes) is stabilized for higher $D$ . b and c, shows that these magnetic states are also observed when DMI becomes larger. Representative configurations with mixed phase, isolated skyrmions and uniform magnetization are also presented.}
\label{Brandao5-1}}
\end{figure*}


Finally, we performed further simulations to build up a phase diagram to explore  the physical parameters needed to stabilize skyrmions at zero magnetic field. The simulated  $D$ $vs$ $M_s$ phase diagram  is shown in Fig. \ref{Brandao5-1}. It  distinguishes three different phases as the ground state. In Fig \ref{Brandao5-1} a, the mixed phase, in which isolated skyrmions and magnetic stripes coexist, is more favorable independently of $M_s$ when $D$ is higher than 0.4 mJm$^{-1}$. A representative configuration of this mixed phase is shown on the bottom of the Fig \ref{Brandao5-1} (a). Between $D$ = 0.2 - 0.4 mJm$^{-1}$ for low values of $M_s$, a narrow region is  observed which isolated skyrmions are stabilized. For lower $D$ values below 0.2 mJm$^{-1}$, the uniform magnetization is the ground state.  Interestingly, increasing  $k$, the  region with stabilized skyrmions is observed in a wider region, see Fig \ref{Brandao5-1} (b). A representative configuration of this stabilized skyrmions is shown on the bottom of the Fig \ref{Brandao5-1} (b).  Besides, it is noted that the uniform magnetization pattern increases while the mixed phase is reduced.  For higher $k$ the isolated skyrmions are stabilized for larger $D$, see Fig \ref{Brandao5-1} (c). The mixed phase is drastically reduced, while the uniform magnetization is mostly observed as the ground state. A representative configuration of the uniform magnetization is shown on the bottom of the Fig \ref{Brandao5-1} (c). The diagram shows how the physical parameters can be tailored in order to achieve the desired magnetic pattern, which is an important mechanism  to designing room-temperature skyrmions based devices.



In conclusion, we investigated the formation of magnetic domains patterns in Pd/Co/Pd multilayers  using MFM images. By thinning  the Co  interlayer we observed a transition from worm-like magnetic domains pattern to isolated skyrmions at room temperature and zero magnetic  field. Stabilized skyrmions were observed in samples in both as-grown and remnant states, and their homogeneity was proved by several MFM images taken over different areas of the sample surface. Comparing the measured and simulated skyrmions size, the quantitative interfacial  Dzyaloshinskii-Moriya interaction iDMI was estimated with $D$ ranging from 0.4 - 0.6 mJm$^{-1}$ as one of the key physical properties on the skyrmions stability. The symmetric multilayers grown on Si substrate by magnetron sputtering are very suitable to host skyrmions in unpatterned samples without any stabilizing field,  providing  a new path towards skyrmions-based room temperature devices. They can be used in areas exceeding 10$\times$10 $\mu$m$^{2}$, and eventually manipulated with ultra-low current density in order to perform devices combined with narrow constrictions without the need to transform magnetic worm-like domains in skyrmions.


\section{Methods}

\subsection{Film deposition and magnetic characterization}

The Pd/Co/Pd multilayers were grown onto Si substrates by magnetron sputtering at room temperature and deposition pressure of 2 mTorr in the argon atmosphere. Two identical samples were  grown in each sputtering process in order to acquire MFM images at zero magnetic field and also magnetization curves. To increase the magnetic contrast for MFM measurements and the signal-to-noise ratio for magnetic hysteresis curves, fifteen repetitions of the trilayers were grown.   X-rays reflectometry (XRR) measurements were performed to verify the Pd and Co thicknesses. The magnetization reversal as a function of magnetic field was acquired by using an alternating gradient field magnetometer (AGFM) and vibrating-sample magnetometer (VSM). 

\subsection{Magnetic Force Microscopy Images}

MFM images were carried out using a Nanosurf FlexAFM microscope. We used MagneticMulti75-G MFM tips from Budget Sensors, they are coated by a cobalt alloy having magnetic moment of roughly 10$^{-16}$ Am$^{2}$ and coercivity of roughly 0.03 T. We operated the MFM measurements at the dynamic force mode with a resonant frequency of about 75 kHz. The images were acquired in the  tip-surface distance of 60 nm. To acquire images in different areas, we moved  the sample over distances of 1 mm in order to confirm the homogeneity of the domain magnetic patterns. 

\subsection{Micromagnetic Modeling}

Micromagnetic simulations were performed using the Mumax$^{3}$ GPU-accelerated program  over an area of 1.5$\times$1.5 $\mu$m$^2$ discretized in cells size of 3$\times$3$\times$3 nm$^3$. The magnetic parameters kept fixed  exchange stiffness  $A_{exch}$ = 15 pJm$^{-3}$ and Gilbert damping $\alpha$ = 0.3. The magnetic anisotropy $k$, magnetization saturation $M_{s}$, and Dzyaloshinskii-Moriya vector $D$ were modified to understand their impact on the skyrmions stability. To obtain the magnetic ground state, the initial magnetization is chosen randomly to seek for the same aleatory distribution of the experimental observed skyrmions. Subsequently, the magnetization is left to relax for 1 $\mu$s, then  by minimizing the  energies involved, reach  an equilibrium condition which represents the skyrmions stability. The long time 1 $\mu$s, to carry out the simulations, was chosen in order to verify if the  skyrmions would persist after being stabilized. No transformation after their stability was observed.   To clarify the influence of the magnetic parameters in the stabilized skyrmions shown in the phase diagram, further simulations were done using negative values of iDMI. Comparable results were obtained regarding the ferromagnetic order and  skyrmions stability.

\section{Acknowledgments}



This work was partially supported by the Brazilian agencies FAPESP (Project No. 2012/51198-2), CAPES and CNPq. J.B., D.A.D. and R.L.S. gratefully acknowledge fellowships provided by CAPES. J.C.C. acknowledges support from CNPq (Project No. 309354/2015-3). The authors also thank the Scientific Computing Group and the XRD2 beamline of Brazilian Synchrotron Light Laboratory (LNLS/CNPEM) for the computer facilities to undertake the micromagnetic simulations and some of the XRR measurements, respectively.

\section{Author Contributions}

J.B, T.J.A.M and J.C.C conceived the project. D.A.D and R.L.S grew the samples and measured the hysteresis curves in the (AGFM) with the support of J.C.D. J.B carried out the MFM images and micromagnetic simulations. J.B, T.J.A.M and J.C.C prepared the manuscript. All authors participated and discussed on the manuscript.


\begin{thebibliography}{10}


\bibitem{jeovani1} Fert, A., Cros, V. and Sampaio, J. Skyrmions on the track. Nat. Nanotechnol. {\bf  8}, 152–156 (2013).

\bibitem{jeovani2} X. Zhang, G. P. Zhao, Hans Fangohr, J. Ping Liu, W. X. Xia, J. Xia and F. J. Morvan. Skyrmion-skyrmion and skyrmion-edge repulsions in skyrmion-based racetrack memory, Sci Rep  {\bf 5}, 7643 (2015).




\bibitem{jeovani6} Krause, S., Wiesendanger, R. Spintronics: Skyrmionics gets hot.  Nature Materials, {\bf 15} (5), pp. 493-494 (2016). 


\bibitem{review} Naoto Nagaosa and  Yoshinori Tokura
 Topological properties and dynamics of magnetic skyrmions. Nat. Nanotechnol. {\bf 8}, 899–911 (2013)


\bibitem{jeovani8} Fert, A., Reyren, N. and  Cros, V. Magnetic skyrmions: advances in physics and potential applications. Nat. Rev. Mat. {\bf 2}, 17031 (2017).

\bibitem{jeovani9} Jiang, W. et al. Skyrmions in magnetic multilayers. Phys. Rep. {\bf 704}, 1–49 (2017).

\bibitem{jeovani9a} Sampaio, J., Cros, V., Rohart, S., Thiaville, A. and  Fert, A. Nucleation, stability and current-induced motion of isolated magnetic skyrmions in nanostructures. Nat. Nanotechnol. {\bf 8}, 839–844 (2013).



\bibitem{jeovani9c} Guoqiang Yu, Pramey Upadhyaya, Qiming Shao, Hao Wu, Gen Yin, Xiang Li, Congli He, Wanjun Jiang, Xiufeng Han, Pedram Khalili Amiri, and Kang L. Wang. Room-Temperature Skyrmion Shift Device for Memory Application, Nano Lett {\bf 17},  261–268 (2017). 


























\bibitem{jeovani19} Dzyaloshinsky, I. A thermodynamic theory of weak ferromagnetism of antiferromagnetics. J. Phys. Chem. Solids {\bf 4}, 241–255 (1958).

\bibitem{jeovani20} Moriya, T. Anisotropic superexchange interaction and weak ferromagnetism. Phys. Rev. {\bf 120}, 91–98 (1960).

\bibitem{jeovani21} A. Thiaville et al. Dynamics of Dzyaloshinskii domain walls in ultrathin magnetic films. Europhys. Lett. {\bf 100}, 57002 (2012).


\bibitem{jeovani7} Wiesendanger, R. Nanoscale magnetic skyrmions in metallic films and multilayers: a new twist for spintronics. Nat. Rev. Mat. {\bf 1}, 16044 (2016).

\bibitem{jeovani22} Yoshimura, Y. et al. Soliton-like magnetic domain wall motion induced by the interfacial Dzyaloshinsky-Moriya interaction. Nat. Phys {\bf 12}, 161 (2016).





\bibitem{jeovani23} Stashkevich, A. A. et al. Experimental study of spin-wave dispersion in Py/Pt film structures in the presence of an interface Dzyaloshinskii-Moriya interaction. Phys. Rev. B {\bf 91}, 214409 (2015).


\bibitem{jeovani24} Yang, H., Thiaville, A., Rohart, S., Fert, A. and Chshiev, M. Anatomy of Dzyaloshinskii-Moriya Interaction at Co/Pt Interfaces. Phys. Rev. Let {\bf 115}, 267210 (2015).

\bibitem{jeovani25} Jaehun Cho, N.-H. et al. Thickness dependence of the interfacial Dzyaloshinsky-Moriya  interaction in inversion symmetry broken systems. Nat Commun {\bf 6}, 7635 (2015).


\bibitem{jeovani26} C. Moreau-Luchaire at al. Additive interfacial chiral interaction in multilayers for stabilization of small individual skyrmions at room temperature.  Nat. Nanotech  {\bf 11},  444–448 (2016). 

\bibitem{jeovani27} Anjan Soumyanarayanan et al. Tunable room-temperature magnetic skyrmions in
Ir/Fe/Co/Pt multilayers. Nat. Mater {\bf 16}, 898–904 (2017). 

\bibitem{chiral} J. C. Chauleau et al. Chirality in Magnetic Multilayers Probed by the Symmetry and the Amplitude of Dichroism in X-Ray Resonant Magnetic Scattering Phys. Rev. Lett. {\bf 120}, 037202 (2018). 


\bibitem{jeovani28}  Gong Chen, Arantzazu Mascaraque, Alpha T. N'Diaye, and Andreas K. Schmid. Room temperature skyrmion ground state stabilized through interlayer exchange coupling,  Appl. Phys. Lett. {\bf 102}, 222405 (2013).

\bibitem{jeovani29} A. Hrabec, J. Sampaio, M. Belmeguenai, I. Gross, R. Weil, S.M. Cherif, A. Stashkevich, V. Jacques, A. Thiaville and S. Rohart, Current-induced skyrmion generation and dynamics in symmetric bilayers, Nat. Comm {\bf 8}, 15765 (2017). 

\bibitem{jeovani30} Seonghoon Woo et al. Observation of room-temperature magnetic skyrmions and their current-driven dynamics in ultrathin metallic ferromagnets.  Nat. Mat {\bf 15}, 501–506 (2016). 


\bibitem{jeovani31} O. Boulle at al. Room-temperature chiral magnetic skyrmions in ultrathin magnetic nanostructures, Nanotech {\bf 11}, 449–454 (2016). 


\bibitem{jeovani32} Pin Ho, Anthony K.C. Tan, S. Goolaup, A.L. Gonzalez Oyarce, M.Raju, L.S. Huang, Anjan Soumyanarayanan, C. Panagopoulos, Sub-100 nm Skyrmions at Zero Magnetic Field in Ir/Fe/Co/Pt Nanostructures, arXiv:1709.04878 (2017). 











\bibitem{jeovani35} William Legrand, et al. Room-Temperature Current-Induced Generation and Motion of sub-100 nm Skyrmions,  Nano Lett {\bf 17}, 2703–2712 (2017).









\bibitem{jeovani39} Fert A. Magnetic and transport properties of metallic multilayers Mater. Sci. Forum, 59–60 (1990), pp. 439-480
CrossRefView Record in Scopus. 

\bibitem{jeovani40} Crepieux A., Lacroix C. Dzyaloshinsky-Moriya interactions induced by symmetry breaking at a surface J. Magn. Magn. Mater {\bf 198}, 341-349 (1998).


\bibitem{jeovani41} Hrabec, A. et al., Measuring and tailoring the Dzyaloshinsky-Moriya  interaction in perpendicularly magnetized thin films. Phys. Rev. B {\bf 90}, 020402 (2014).



\bibitem{jeovani42} Kim, D.O. et al. Asymmetric magnetic proximity effect in a Pd/Co/Pd trilayer system. Sci. Rep. {\bf 6}, 25391 (2016).

\bibitem{jeovani43} Ryu, K.S., Yang, S.-H., Thomas, L. and Parkin, S. S. P. Chiral spin torque arising from proximity-induced magnetization. Nat. Commun. {\bf 5}, 3910 (2014).


\bibitem{jeovani34}  Shawn D. Pollard, Joseph A. Garlow, Jiawei Yu, Zhen Wang, Yimei Zhu and Hyunsoo Yang. Observation of stable N\'{e}el skyrmions in cobalt/palladium multilayers with Lorentz transmission electron microscopy, Nat. Comm  {\bf 8},  14761 (2017). 

\bibitem{skyrmions200nm} A. Kolesnikov et al. 	arXiv:1709.01229 (2017). 


\bibitem{size} N. Romming et al. Field-Dependent Size and Shape of Single Magnetic Skyrmions. Phys. Rev. Lett. {\bf 114}, 177203 (2015). 


\bibitem{large} Jiang, W. et al. Blowing magnetic skyrmion bubbles. Science {\bf349}, 283 – 286 (2015).

\bibitem{Mumax}Arne Vansteenkiste et al. The design and verification of mumax, AIP Advances 4, 107133 (2014).









\bibitem{jeovani45} J. M. Shaw, H. T. Nembach, T. J. Silva, S. E. Russek, R. Geiss, C. Jones,N. Clark, T. Leo, and D. J. Smith. Effect of microstructure on magnetic properties and anisotropy distributions in Co/Pd thin films and nanostructures Phys. Rev. B {\bf 80}, 184419 (2009).



\bibitem{jeovani46} Z. Liu, R. Brandt, O. Hellwig, S. Florez, T. Thomson, B. Terris, and H.Schmidt,Thickness dependent magnetization dynamics of perpendicular anisotropy Co/Pd multilayer films. J. Magn. Magn. Mater. {\bf 323}, 1623 (2011).

\bibitem{jeovani47} Y. Kachlon et al., Extracting magnetic anisotropy energies in Co/Pd multilayers via refinement analysis of the full magnetoresistance curves. J. Appl. Phys. {\bf 115}, 173911 (2014).














 










 





















\end{thebibliography}
\end{document}